\documentclass[runningheads]{llncs}

\usepackage[T1]{fontenc}
\def\doi#1{\href{https://doi.org/\detokenize{#1}}{\url{https://doi.org/\detokenize{#1}}}}
\usepackage{graphicx}
\usepackage{listings}
\usepackage[misc]{ifsym} 
\usepackage{amsmath}
\usepackage{amssymb}
\usepackage{booktabs}
\usepackage{multirow}
\usepackage{url}
\usepackage[dvipsnames]{xcolor}
\begin{document}
\title{Graph Neural Networks with Dynamic and Static Representations for Social Recommendation}
\titlerunning{GNNs with Dynamic and Static Representations for Social Recommendation}

\author{Junfa Lin \and
Siyuan Chen \and
Jiahai Wang \inst{\textrm{\Letter}}}
\authorrunning{J. Lin et al.}
\institute{School of Computer Science and Engineering, Sun Yat-sen University, Guangzhou, China\\
\email{\{linjf26,chensy47\}@mail2.sysu.edu.cn},
\email{wangjiah@mail.sysu.edu.cn}}
\maketitle              % typeset the header of the contribution
\begin{abstract}
    Recommender systems based on graph neural networks receive increasing research interest due to their excellent ability to learn a variety of side information including social networks. However, previous works usually focus on modeling users, not much attention is paid to items. Moreover, the possible changes in the attraction of items over time, which is like the dynamic interest of users are rarely considered, and neither do the correlations among items. To overcome these limitations, this paper proposes graph neural networks with dynamic and static representations for social recommendation (GNN-DSR), which considers both dynamic and static representations of users and items and incorporates their relational influence.
    GNN-DSR models the short-term dynamic and long-term static interactional representations of the user's interest and the item's attraction, respectively. Furthermore, the attention mechanism is used to aggregate the social influence of users on the target user and the correlative items' influence on a given item. 
    The final latent factors of user and item are combined to make a prediction. Experiments on three real-world recommender system datasets validate the effectiveness of GNN-DSR.
\end{abstract}
\section{Introduction}
% Introduction to background and limitations in previous recommendation methods.
% The motivation in this paper is presented and illustrated with Figure \ref{fig:session}

% With the rapid development of online communities, the amount of information people are faced with in their daily lives is constantly increasing. Recommender systems, which help users to find items of potential interests, mitigate the information overload problem, and have extensive applications \cite{zhang2019deep}.
Recommender systems, which help users to find their potential interests of items, mitigate the information overload problem, and have extensive applications \cite{zhang2019deep}.
According to the social homophily hypothesis \cite{mcpherson2001birds}, two closely related users in the social network may have common or similar interests. Social recommender systems are increasing rapidly by considering social influences \cite{tang2013social}. In addition to these explicit user social influences, the correlation between two items is also important. It provides extra information that may describe the items since items are likely to be similar or related \cite{sarwar2001item}. 
% They alleviate the problem of data sparsity and further improve recommendation performance. 
At the same time, most of the information including social networks essentially has a graph structure, and graph neural networks (GNNs) have a powerful capability in graph representation learning, thus the field of utilizing GNNs in recommender systems is flourishing \cite{wu2020graph}.

% In addition to these explicit user social relationships in the social recommendation, the correlation between two items is also important. It provides extra information that may be used to describe the items since items are likely to be similar or related \cite{sarwar2001item}. 
% For example, a user who has purchased an Apple's new iPhone is likely to consider purchasing the Apple AirPods as they are highly correlative. 
% Similar to user social graphs, these correlations among items can be represented as item correlative graphs. Therefore, to enhance the learning of item representations in the social recommendation, it is desirable to consider the correlations between items \cite{fan2020graph}.

% First, many of the social recommenders only capture users' general preferences by static user-item interactions \cite{fan2019graph,fan2020graph,ma2011recommender,wang2017learning,zhao2014leveraging}. In practical applications, user information and activities are continuously recorded on most web platforms. Dynamic behaviors will reflect changes in user interests over time. 
% Second, works that have considered dynamic user behaviors, such as session-based recommendations \cite{hidasi2015session,li2017neural,wu2019session,yuan2019simple}, usually consider implicit feedback between a user and an item without relational influence. 

However, previous works usually focus on modeling users' behaviors with little attention to items. For the modeling of items, they have rarely modeled the possible changes in the attraction of items over time and the correlations among items, which are like the users' interests and their social influence, respectively. 

\begin{figure}[t]
    \centering
    \includegraphics[width=0.8\linewidth]{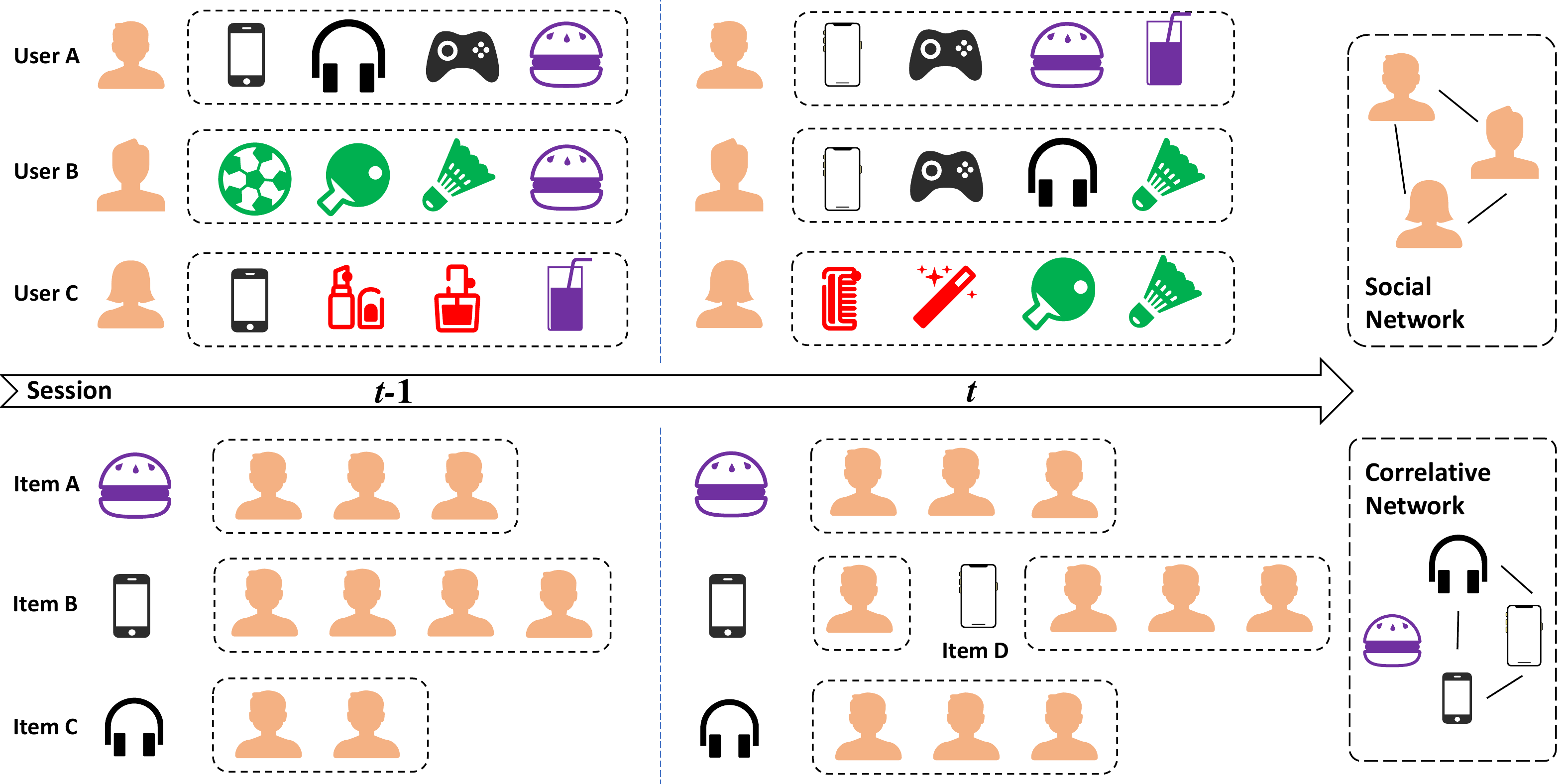}
    \caption{An example of possible changes in the interest of users and the attraction of items over time.}
    \label{fig:session}
  \end{figure}

% \textbf{Motivating Example}. 
Figure \ref{fig:session} presents a motivating example that the interest of users and the attraction of items may change over time and be influenced by their relationships. On the user side, user B is a common friend of user A and user C, but their interests may not be identical. At time $t-1$, the only common interest of the three users is in food; however, by time $t$, user B may be influenced by user A to develop an interest in digital products, while user C may be influenced by user B to develop a new interest in sports. Thus, a dynamic influence of friends on the user can facilitate the learning of the target user's current interest representations. At the same time, users may also have their long-term static interests, e.g. user A may always be interested in digital products and user C may always be interested in beauty. 
On the item side, items may exhibit similar patterns to users. The attraction of items will also change with time. For example, some digital products will gradually become outdated due to updates, even if they are popular products in the current period. Item B, as shown in Figure \ref{fig:session}, is popular with users at time $t-1$. But at time $t$, the emergence of a new generation of products like item D, suppresses the attraction of item B. However, there are also items such as item A and item C that are always popular with users because of their excellent quality. Therefore, in addition to considering the correlations among items, the attraction of items over time should also be considered. Collectively, interactional representations can be referred to the user's interest and the attraction of item. Relational influences refer to the social influence of the user and the correlative influence of the item.

% \textbf{Methodologies}. 
To overcome these limitations and to focus on the attraction of items over time, we propose graph neural networks with dynamic and static representations of users and items for social recommendation (GNN-DSR).
% , which considers both users' and items' interactional representations and relational influences. 
GNN-DSR consists of two main components: interaction aggregation and relational graph aggregation. For both users and items in the interaction aggregation, the recurrent neural networks (RNNs) are used to model the short-term dynamic representations and the graph attention mechanism is utilized on historical user-item interactions to model the long-term static representations.
In the relational graph aggregation, the influences from the user-user graph and the item-item graph that are termed as the relational graphs, are aggregated via the graph attention mechanism over users' or items' representations. 
GNN-DSR then combines the interactional representations and relational influences through a multi-layer perceptron (MLP) for the rating prediction. 
% In the item modeling part, a similar structure with user modeling is designed, using an LSTM to model the short-term attraction of items and using a GAT to model the long-term attraction representations. The correlative influence of related items in the item correlative graph is aggregated through a GAT and is then combined with attraction representations through an MLP for item latent factor modeling.
% The obtained user and item latent factors are concatenated for the rating prediction. 
The contributions of this paper can be summarized as:
\begin{itemize}
  \item We propose an effective framework called GNN-DSR for social recommendation, which models both dynamic and static representations of users and items, as well as incorporating their relational influence.
  \item On the observation that items have the dynamic and static attraction like the user's interests, our method captures the dynamic representations through RNNs and introduces the graph attention mechanism to model static representations. GNN-DSR also captures the relational influences in both user and item domains through user social and item correlative graphs, respectively.
  \item Extensive experiments are conducted on three real-world data sets. The experimental results verify the effectiveness of the proposed model.
\end{itemize}

\section{Related Work}
In this section, we briefly review two kinds of related work including session-based recommendation and social recommendation.
% Social recommendation, on the other hand, emerged to exploit the social relationships among users to predict their unknown preferences.

\textbf{Session-based Recommendation.}
The session-based recommendation is a sequential task of recommender systems that predicts users’ next preferences based on their most recent activities \cite{wang2021survey}. Recurrent neural networks have proved effective for the sequential problem. 
Hidasi et al. \cite{hidasi2015session} first proposed an RNN-based approach for session-based recommendation. 
% Tan et al. \cite{tan2016improved} further applied some popular techniques to improve RNNs'
% performance for session-based recommendations.
Li et al. \cite{li2017neural} captured both the user’s local and global interests by utilizing an attention mechanism into RNNs. 
Wang et al. \cite{wang2019modeling} explicitly addressed two item-specific temporal dynamics: short-term effect and lifetime effect. 
Recently, GNNs have achieved excellent performance in various tasks including session-based recommendations. 
% Wu et al. \cite{wu2019session} constructed a directed session graph and used a GNN to obtain complex transitions of items for sequence tasks.
% Xu et al. \cite{xu2019graph} proposed a graph contextualized self-attention model, which utilizes both GNN and self-attention mechanism to capture more interaction between items in the sequence.
Zhang et al. \cite{zhang2020personalized} used the attention mechanism to explicitly model the effect of the user’s historical sessions on the current session. 
Wang et al. \cite{wang2020next} developed a novel next-item recommendation framework empowered by sequential hypergraphs. 

This line of research mainly focuses on the user-item sequential interactions and some methods \cite{wang2019modeling,wang2020next} further consider the dynamic changes of items. However, they do not consider the relational influences among users or items. Our method incorporates both sequential information and relational influence into learning representations for users and items.

\textbf{Social Recommendation.}
On account of studies for social effects, such as homophily \cite{mcpherson2001birds} and influence \cite{marsden1993network}, many previous studies attempted to leverage social networks to improve the recommendation results.
Ma et al. \cite{ma2011recommender} elaborated how social network information can benefit recommender systems. 
Zhao et al. \cite{zhao2014leveraging} used social connections to better estimate users’ rankings of products. 
% Xiao et al. \cite{xiao2017learning} adopted transfer learning to model user-item interactions and social relationships simultaneously. 
Wang et al. \cite{wang2017learning} distinguished and personalized the importance of different types of ties in the social network.
As the development of GNNs, many social recommendations based on GNNs have also been proposed.
Fan et al. \cite{fan2019graph} proposed a GNN to jointly capture representations of the user-item graph and user-user social graph for social recommendations.
Wu et al. \cite{wu2020diffnet++} modeled user influence diffusion and interest diffusion by reformulating social recommendations as heterogeneous graphs with social network and interest network.
% Lu et al. \cite{lu2020social} proposed a novel social influence attention neural network for the Friend Enhanced Recommendation (FER) problem.
% These methods only capture static information from user-item interactions without considering the dynamic information.
Wu et al. \cite{wu2019dual} proposed a dual graph attention network to collaboratively learn representations for two-fold social effects.
% , by modeling a user-specific attention weight and a context-aware attention weight. 
% Fan et al. \cite{fan2020graph} 
Song et al. \cite{song2019session} modeled dynamic user interests with an RNN and context-dependent social influence with a graph attention network.
Gu et al. \cite{gu2021enhancing} incorporated item graph embedding and contextual social modeling into the recommendation task. 
Chen et al. \cite{chen2021efficient} integrated the knowledge from social networks to learn user and item representations via a heterogeneous GNN.

However, these methods ignore the possible dynamic attraction, as well as the correlations, of the items. 
Due to the benefits of GNNs, we also adopt them in our work. An item correlative graph is constructed in our framework together with the social graph for the rating prediction task.

\section{Problem Definition}

Let $U = \{u_1,\dots, u_n \} $ and $V = \{v_1, \dots, v_m\} $ denote the set of users and items, respectively, where $n$ is the number of users and $m$ is the number of items. $\mathbf{R} \in \mathbb{R} ^{n \times m}$ is the user-item interaction graph. $r_{ij}$ represents the rating value given by user $u_i$ on item $v_j$, and $r_{ij}=0$ if $u_i$ has not rated $v_j$ or it is an unknown interaction. 
Generally, $\mathbf{R}$ is sparse, which needs some side information to alleviate. 
$G_U=(U,E_U)$ is defined as the user social graph, where $E_U$ is the set of edges connecting users.

The set of $u_i$'s friends is denoted as $\mathcal{N}^{uu}_i$, with the superscript `$uu$' representing an undirected edge between two users. $\mathcal{N}^{uv}_i$ is defined as the set of items with which $u_i$ has interacted, while the superscript `$uv$' indicates a directed edge from a user to an item. Meanwhile, the item correlative graph is denoted as $G_V=(V,E_V)$, where $E_V$ is the set of edges connecting related items. $\mathcal{N}^{vv}_j$ denotes the set of items that are correlative with $v_j$. $\mathcal{N}^{vu}_j$ is defined as the set of users who have interacted with $v_j$.
% Users click or rate items within a session. 
Each interaction is recorded as a triple $(u_i,v_j,\tau)$, with $\tau$ as the time stamp. Let $\mathcal{T}^v_i$ be an ascending sequence of all time stamps of $u_i$'s interactions, then the item sequence that $u_i$ has interacted with can be written as $S^v_i=(v_{j(\tau)})_{\tau\in\mathcal{T}^v_i}$. Similarly, let $\mathcal{T}^u_j$ denote an ascending sequence of all time stamps that users interact with $v_j$, then the corresponding user sequence can be written as $S^u_j=(u_{i(\tau)})_{\tau\in\mathcal{T}^u_j}$.

% \textbf{Problem Formulation.} 
The social recommendation problem can be defined as \cite{tang2013social}: given the observed user-item interaction $\mathbf{R}$ and the social relationship $G_U$, the recommender should predict the users' unknown interactions in $\mathbf{R}$, i.e., the rating value of unobserved item $v_j$ user $u_i$ will score (for explicit feedback) or the probability of an unobserved candidate item $v_j$ user $u_i$ will click on (for implicit feedback).

\section{The Proposed Framework}

\begin{figure}[t]
    \includegraphics[width=1.01\textwidth]{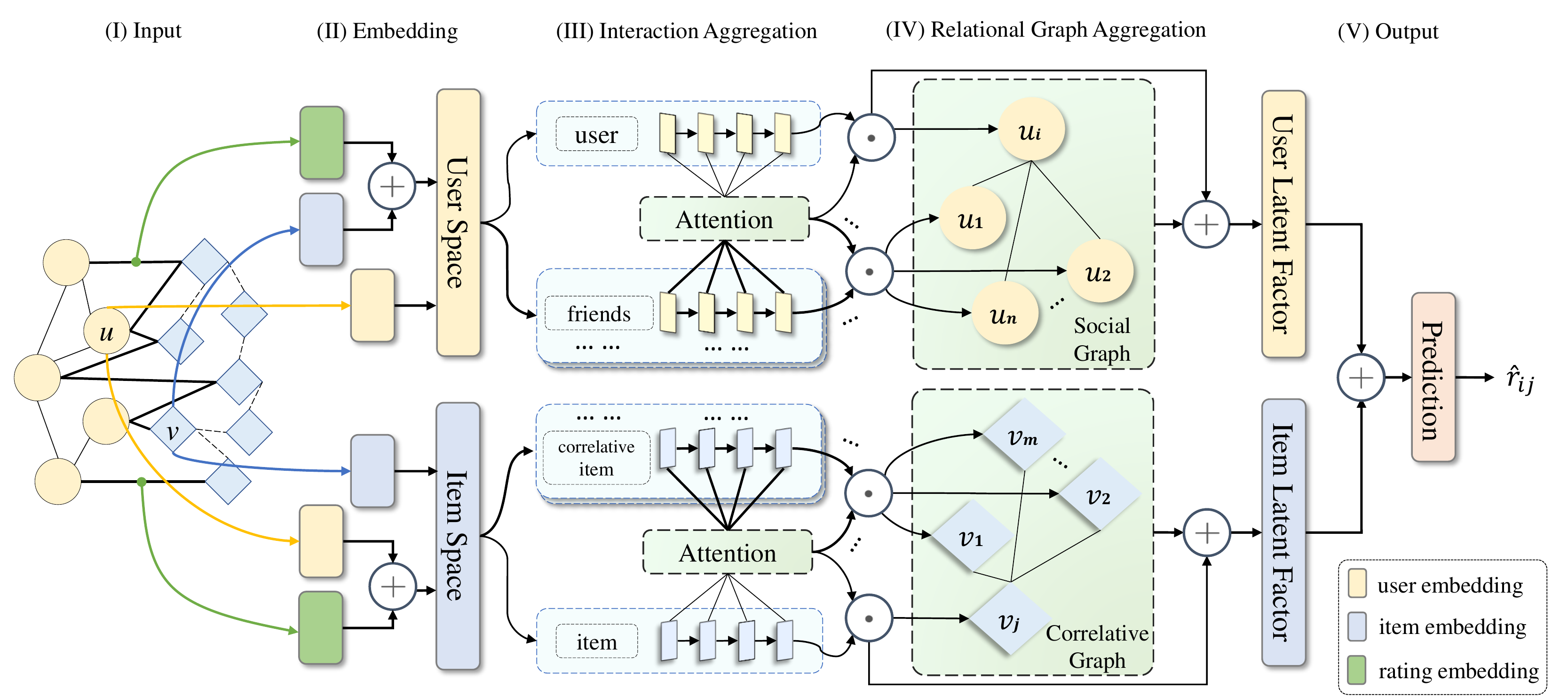}
    \caption{Overview of GNN-DSR. From the left to the right, GNN-DSR can be divided into five layers: 1) Input Layer, which processes the raw data and feeds them into the GNN-DSR. 2) Embedding Layer, which transforms the preprocessed data into the user vector space and item vector space, respectively. 3) Interaction Aggregation, which models both dynamic and static representations of users and items, respectively. 4) Relational Graph Aggregation, which aggregates the impacts of neighboring users or items, respectively. 5) Output Layer, which uses an MLP to make a rating prediction.}
    \label{fig:model}
\end{figure}


In this section, our framework, called GNN-DSR, will be introduced in detail.
The overview of GNN-DSR is shown in Figure \ref{fig:model}. 
The input layer preprocesses the raw data and constructs the item correlative graph. Then, the embeddings of user and item are transformed from the preprocessed data. These embeddings are fed to the core components of our method, interaction aggregation and relational graph aggregation, which captures the user-item interactions and the relational influences, respectively.
The final output layer integrates the latent factor of user and item to predict the ratings. 

\subsection{Input Preprocessing}
\label{Input}
Our work requires the user-item interaction graph $\mathbf{R}$, the user social graph $G_U$, and the item correlative graph $G_V$ as the inputs. 
% Therefore, the main task of the input layer is to preprocess the original raw data, focusing on the item correlative graph construction. 
Since most datasets do not have prior information that explicitly represents the correlations between items, a natural approach is to connect two items with high similarities \cite{deshpande2004item}. Cosine similarity is used to measure the similarity between two items \cite{sarwar2001item}. For any $v_j$, the $j$-th column of $\mathbf{R}$ records the ratings from all users that have interacted (clicked or rated) with $v_j$. The rating vectors of $v_j$ and $v_k$ are used to calculate the similarity:

\begin{equation}
  \label{eq:cos}
  \text{sim}(j,k)=\frac{\mathbf{R}(\cdot,j)^T\mathbf{R}(\cdot,k)}{\left\lVert \mathbf{R}(\cdot,j) \right\rVert_2\left\lVert \mathbf{R}(\cdot,k) \right\rVert_2} .
\end{equation}

Finally, based on the similarity of items, the top-$k$ similar items are extracted as the correlative neighbors of each item to construct the item correlative graph $G_V$ before the item embedding. 
% The number of edges in the graph is $k\times|V|$.

\subsection{User and Item Embedding}

% The original input for every user and item is a high-dimensional one-hot vector, while the embedding operation maps it to a low-dimensional representation. 
Each user can be characterized by her/his own embedding and the items she/he has rated \cite{koren2008factorization}, while an item can be represented as its embedding and the ratings given by users. The user-item interaction graph $\mathbf{R}$ not only contains the interactions between the user and the item, but it also contains the user's ratings or opinions (denoted by $r$) of the item. These ratings of items not only reveal the user's preferences and interests in the items but also reflect the attraction of the item to users, thus helping to model the user's and item's latent factors. Typically, the rating $r_{ij}$ takes a discrete value. For example, each $r_{ij} \in \{1,2,3,4,5\}$ in a 5-star rating system. Inspired by GraphRec \cite{fan2019graph}, our method embeds each rating to a $d$ dimensional vector and uses $\mathbf{e}_{ij}$ to represent the embedding of $r_{ij}$.

Let $\mathbf{p}_i, \mathbf{q}_j\in \mathbb{R}^d$ be the embeddings of user $u_i$ and item $v_j$, respectively. For the interaction between $u_i$ and $v_j$ with rating $r_{ij}$, the interaction embeddings of $u_i$ and $v_j$ are computed as follows:

\begin{equation}
    \label{eq:xij}
    \mathbf{x}_{i\leftarrow j} = g^{uv}([\mathbf{e}_{ij},\mathbf{q}_j]), \quad
    \mathbf{y}_{j\leftarrow i} = g^{vu}([\mathbf{e}_{ij}, \mathbf{p}_i]),
  \end{equation}
where $\mathbf{x}_{i\leftarrow j}$ and $\mathbf{y}_{j\leftarrow i}$ denotes the interaction embedding from $v_j$ to $u_i$, and from $u_i$ to $v_j$, respectively. $[\cdot,\cdot]$ denotes the concatenation operation of two vectors. Both $g^{uv}$ and $g^{vu}$ are two-layer perceptrons.

% Similar to user embedding, each item can be represented by an embedding of the item and the user who interacted with it (called user-based item embedding).
% For the rating $r_{ij}$ given to $v_j$ by $u_i$, the item interaction embedding is denoted as $\mathbf{y}_{j\leftarrow i}$, which is obtained from the user embedding $\mathbf{p}_i$ and the rating embedding $\mathbf{e}_{ij}$ via an MLP:
% \begin{equation}
%   \label{eq:yji}
%   \mathbf{y}_{j\leftarrow i} = \mathbf{MLP}^{vu}([\mathbf{e}_{ij}, \mathbf{p}_i ]).
% \end{equation}
% {\color{orange}W e then combine the item embedding $\mathbf{q}_j$ of item $v_j$ and the item interaction embedding $\mathbf{y}_{j\leftarrow i}$ as an item vector for the next component learning item latent factor.}

\subsection{Interaction Aggregation}
For each user and item, our method is designed to capture the short-term dynamic interests and the long-term static interests of the user, and the item's attraction of the dynamic and static one, respectively. 

\subsubsection{Short-term Dynamic Representation.}
%RNN-based users' interests representation
After obtaining the user's interaction embedding $\mathbf{x}_{i\leftarrow j}$, the items that $u_i$ interacts within all sequences can be written as $\mathbf{X}(i)=(\mathbf{x}_{i\leftarrow j(\tau)})_{\tau\in\mathcal{T}^v_i}$, and so does the user interaction sequence of item, $\mathbf{Y}(j)=(\mathbf{y}_{j\leftarrow i(\tau)})_{\tau\in\mathcal{T}^u_j}$. To capture the dynamic representations, RNNs are used to model the sequence, since RNNs have good modeling capabilities for sequential data.
Specifically, the long short-term memory (LSTM) \cite{hochreiter1997long} is applied in our method:
\begin{equation}
  \label{eq:h_iS}
  \mathbf{h}_{i}^S = \mathbf{LSTM}(\mathbf{X}(i)), \quad
  \mathbf{h}_{j}^S = \mathbf{LSTM}(\mathbf{Y}(j)),
\end{equation}
where $\mathbf{h}_{i}^S$ and $\mathbf{h}_{j}^S$ denotes the dynamic representation of the user's interest, and of the item's attraction, respectively.
Since it is intended for dynamic interests or attraction, our method only needs the output of the last hidden layer of the LSTM. For long item sequences, the lengths are truncated to a fixed value to reduce the computational cost.

% \subsubsection{User Interests Modeling}
% For each user, our method needs to capture her/his short-term dynamic interests and long-term static interests for items.

\subsubsection{Long-term Static Representation.}
%Attention
% It is unreliable to rely on the user's dynamic preferences alone, so the user's long-term static interests also need to be modeled. 
Regarding each interaction embedding $\mathbf{x}_{i\leftarrow j}$ or $\mathbf{y}_{j\leftarrow i}$ as an edge representation of the user-item graph, our method can aggregate the edge representations via an attention mechanism, as follows:
\begin{equation}
  \label{eq:h_iL}
  \mathbf{h}_{i}^L = \sigma (\mathbf{W}_0^{uv} \sum_{j\in \mathcal{N}^{uv}_i}  \alpha_{ij} \mathbf{x}_{i\leftarrow j} + \mathbf{b}_0^{uv}), \quad
  \mathbf{h}_{j}^L = \sigma (\mathbf{W}_0^{vu} \sum_{i\in \mathcal{N}^{vu}_j}  \alpha_{ji} \mathbf{y}_{j\leftarrow i} + \mathbf{b}_0^{vu}),
\end{equation}
where $\mathbf{h}_{i}^L$ is the static interest representation of the target user $u_i$, and $\mathbf{h}_{j}^L$ is the static attraction of the item.
$\sigma$ is a nonlinear activation function, and $\mathbf{W}_0$ and $\mathbf{b}_0$ are the weight and bias of the network. Different from the self-attention in graph attention networks (GAT) \cite{velivckovic2017graph} that ignore the edge features, $\alpha_{ij}$ and $\alpha_{ji}$ represent the learned attentive propagation weights over the central node representations $(\mathbf{p}_i$, $\mathbf{q}_j)$ and the edge representations $(\mathbf{x}_{i\leftarrow j}$, $\mathbf{y}_{j\leftarrow i})$. Formally, taking $\alpha_{ij}$ as an example, they are calculated as:
\begin{equation}
  \label{eq:attention_xij}
  \alpha_{ij} = \frac{\exp (\mathbf{W}_2^{uv} \cdot \sigma(\mathbf{W}_1^{uv} \cdot [\mathbf{p}_i,\mathbf{x}_{i\leftarrow j}  ] + \mathbf{b}^{uv}_1) + \mathbf{b}^{uv}_2) }  {\sum _{j \in \mathcal{N}^{uv}_i} \exp (\mathbf{W}_2^{uv} \cdot \sigma(\mathbf{W}_1^{uv} \cdot [\mathbf{p}_i,\mathbf{x}_{i\leftarrow j}  ] + \mathbf{b}^{uv}_1) + \mathbf{b}^{uv}_2) } ,
\end{equation}
where $(\mathbf{W}_1,\mathbf{b}_1)$ and $(\mathbf{W}_2,\mathbf{b}_2)$ are the weights and biases of the first and second layers of the attention network, respectively.

% In order to obtain higher-level features of each pair of $(u_i,v_j)$ interaction, our method needs at least one learnable linear transformation \cite{velivckovic2017graph}. Here a two-layer neural network is used, combining the user embedding $\mathbf{p}_i$ and the user interaction embedding $\mathbf{x}_{i\leftarrow j}$ to parameterize the attention weights as follows:

% \begin{equation}
%   \label{eq:a_xij}
%   \alpha^*_{ij} = \mathbf{W}_2^{uv} \cdot \sigma(\mathbf{W}_1^{uv} \cdot [\mathbf{p}_i,\mathbf{x}_{i\leftarrow j}  ] + \mathbf{b}^{uv}_1) + \mathbf{b}^{uv}_2,
% \end{equation}
% where $(\mathbf{W}^{uv}_1,\mathbf{b}^{uv}_1)$ and $(\mathbf{W}^{uv}_2,\mathbf{b}^{uv}_2)$ are the weights and biases of the first and second layers of the network, respectively. The attention weights are then normalized using the softmax function to obtain the final attention weights as follows:

For simpliciy, Eqs.~(\ref{eq:h_iL})-(\ref{eq:attention_xij}) can be written in a compact form,
\begin{equation}
    \label{eq:gat_uv}
    \mathbf{h}_{i}^L = f^{uv}\left(\mathbf{p}_i,\left\{\mathbf{x}_{i\leftarrow j}:j\in\mathcal{N}_i^{uv}\right\}\right), \quad
    \mathbf{h}_{j}^L = f^{vu}\left(\mathbf{q}_j,\left\{\mathbf{y}_{j\leftarrow i}:i\in\mathcal{N}_j^{vu}\right\}\right).
\end{equation}

\subsubsection{Interactional Representation.}
The dynamic representations and static representations are directly combined to get the interest representation $\mathbf{h}_{i}^I$ of user and the attraction representation $\mathbf{h}_{j}^A$ of item via Hadamard product, i.e.,
\begin{equation}
  \label{eq:h_iU}
  \mathbf{h}_{i}^I = \mathbf{h}_{i}^S \odot  \mathbf{h}_{i}^L, \quad
  \mathbf{h}_{j}^A = \mathbf{h}_{j}^S \odot  \mathbf{h}_{j}^L.
\end{equation}

Similarly, for a target user $u_i$'s friends $u_o, o\in \mathcal{N}^{uu}_i$ in the social graph or an item $v_k, k \in \mathcal{N}^{vv}_j$ related to $v_j$ in the item correlative graph, we can use the aforementioned method, summarized in Eq.~(\ref{eq:h_iU}), to obtain their interest representations $\mathbf{h}_{o}^I$ or attraction representation $\mathbf{h}_{k}^A$, respectively.

\subsection{Relational Graph Aggregation}
\subsubsection{Social Aggregation for User.}
% According to the social homogeneity hypothesis \cite{mcpherson2001birds}, users' interests may be similar to or influenced by their directly connected friends. 
The social information should be combined to further model the latent factors of users. Meanwhile, users may have more similar interests in strong relationships than in weak relationships. Modeling users' latent factors under social network information should take into account the heterogeneity of social relationship strength. 
Therefore, the social influence $\mathbf{h}_{i}^N$ of user is aggregated by the attention mechanism from the social graph, which is calculated as follows:

% Therefore, a GAT is also introduced to obtain the contribution of different social friends to user influence and aggregate information about them.

% The interest representations $\mathbf{h}_i^I$ and $\mathbf{h}_o^I$ are combined to model the social influence $\mathbf{h}_i^N$ of user via the user social graph. The GAT is used to aggregate the adjacent users' interest representations from the social graph, which is calculated as follows.

% \begin{equation}
%   \label{eq:socialagg}
%   \mathbf{h}_i^N = \sigma \left(\mathbf{W}_0^{uu} \cdot \sum_{o\in \mathcal{N}_i^{uu}}  \beta_{io} \mathbf{h}_o^I + \mathbf{b}_0^{uu}\right),
% \end{equation}
% where $\beta_{io}$ is the social attention weight of the adjacent users' interest representations. The attention mechanism is likewise modeling through a two-layer neural network to extract friends who have significant influence on the target user $u_i$. Adjacent user interest representations $\mathbf{h}_o^I$ and the target user's embedding $\mathbf{p}_i$ are combined for the $\beta_{io}$ modeling, as follows:

% \begin{equation}
%   \label{eq:beta}
%   \begin{aligned}
%   \beta^*_{io} & = \mathbf{W}_2^{uu} \cdot \sigma(\mathbf{W}_1^{uu} \cdot [\mathbf{h}_o^I,\mathbf{p}_i ] + \mathbf{b}_1^{uu}) + \mathbf{b}^{uu}_2    ,\\
%   \beta_{io} &. = \frac{\exp (\beta^*_{io}) }  {\sum _{o \in \mathcal{N}_i^{uu}} \exp (\beta^*_{io}) }.
%   \end{aligned}
% \end{equation}

% For simpliciy, Eqs.~(\ref{eq:socialagg})-(\ref{eq:beta}) can be summarized as
\begin{equation}
    \label{eq:h_iN}
    \mathbf{h}_{i}^N = f^{uu}\left(\mathbf{p}_i,\left\{\mathbf{h}^I_{o}:o\in\mathcal{N}_i^{uu}\right\}\right).
\end{equation}

\subsubsection{Correlative Aggregation for Item.}
Since items are not independent, there are likely other similar or correlative items. To further enrich the item latent factors from the item correlative graph $G_V$ is reasonable. Similar to the user's one, item correlative representation $\mathbf{h}_{j}^N$ is calculated as follows:
% Similar to the user's social aggregation, item correlative aggregation takes the adjacent items $v_k$ interaction embeddings $\mathbf{z}_k^A$ of $v_j$ to form a new item correlative graph. The nodes of the newly constructed graph are items interaction embeddings, and then the GAT is used to distinguish the strength of the influence from different items. Therefore our method can model the item correlative representation $\mathbf{z}_j^N$, which is calculated as follows:

\begin{equation}
  \label{eq:h_jN}
  \mathbf{h}_{j}^N = f^{vv}\left(\mathbf{q}_j,\left\{\mathbf{h}_{k}^A:\,k\in\mathcal{N}_j^{vv}\right\}\right).
\end{equation}

\subsection{Output Layer}
\label{prediction}

% \subsubsection{User Modeling}
% \label{User modeling}
% User modeling is aimed to model user latent factor, which is denoted as $\mathbf{h}_i$ for user $u_i$. To effectively combine information from the user-item graph and the social graph, two kinds of information aggregation are used to model latent factors. Interests aggregation is used to capture user interest latent factors $\mathbf{h}_i^I$ from $\mathbf{R}$. Social aggregation aims to obtain the social influence $\mathbf{h}_i^N$ of users from their social graph. These two factors are then combined to form the user latent factor $\mathbf{h}_i$. 

\subsubsection{Latent Factor of User and Item.}
To better model the latent factors of user and item, we need to consider the interaction-based representations and the relational-based representations together, since both the user-item interaction and the relational graph provide different perspectives of information about users and items. 
% These two representations are combined to as the final user latent factor through a standard $\mathbf{MLP}$, where the user interest representation $\mathbf{h}_i^I$ and the user social representation $\mathbf{h}_i^N$ are concatenated before feeding into $\mathbf{MLP}$. 
Therefore, the latent factor $\mathbf{h}_i^u$ of $u_i$ and $\mathbf{h}_j^v$ of $v_j$ are defined as

\begin{equation}
  \label{eq:hi}
  \mathbf{h}_i^u = g^{uu}\left(\left[ \mathbf{h}_{i}^I,\mathbf{h}_{i}^N \right]\right), \quad
  \mathbf{h}_j^v = g^{vv}\left(\left[ \mathbf{h}_{j}^A,\mathbf{h}_{j}^N \right]\right).
\end{equation}

% \subsubsection{Item Modeling}
% \label{Item Modeling}
% Item modeling aims to model the latent factor $\mathbf{z}_j$ of an item $v_j$. As introduced before, items are associated with both users in the user-item interaction graph and correlative items in the item correlative graph. To combine these two graphs efficiently, GNN-DSR adopts a similar approach like user modeling, using two kinds of aggregation to model two different item representations from these two graphs, as shown in Figure \ref{fig:model}. Attraction aggregation captures item attraction representations $\mathbf{z}_j^A$ from the user-item interaction. Correlative aggregation obtains item correlative representations $\mathbf{z}_j^N$ based on item correlationships from the item correlative graph. These two representations are then combined to model the final item latent factor $\mathbf{z}_j$.

% \subsubsection{Item Latent Factor}
% The item attraction representation $\mathbf{z}_j^A$ and the correlative representation $\mathbf{z}_j^N$ are combined by an $\mathbf{MLP}$ to model the final item latent factor $\mathbf{z}_j$:
% \begin{equation}
%   \label{eq:latentfactor}
%   \mathbf{z}_j = \mathbf{MLP}^{vv}\left(\left[\mathbf{z}_j^A,\mathbf{z}_j^N\right]\right).
% \end{equation}

\subsubsection{Prediction}
Our method is mainly applied to the recommendation task of rating prediction, which calculate the predicted rating $\hat{r}_{ij}$ as follows:
\begin{equation}
  \hat{r} _{ij} = g_{\text{output}}\left(\left[\mathbf{h}_i^u, \mathbf{h}_j^v \right]\right),
\end{equation}
where $g_{\text{output}}$ is a multilayer perceptron with three layers.

\subsection{Training}
\label{training}
% To train the model parameters, an optimizable objective function is a need.
Let $\mathcal{P} = \left\{ (i, j) : r_{ij} \neq 0 \right\}$ be the set of known ratings in the dataset. The mean squared error is used for training the model:
\begin{equation}
  \label{eq:loss}
  \text{MSELoss} = \frac{1}{2 \left\lvert \mathcal{P} \right\rvert } \sum_{(i,j) \in
   \mathcal{P} } \left( \hat{r} _{ij} - r_{ij}\right)^2  ,
\end{equation}
where $r_{ij}$ is the true rating score. GNN-DSR is optimized using gradient descent. To alleviate the overfitting problem, the Dropout \cite{srivastava2014dropout} is applied to our work.

To allow ranking predictions of items, some modifications to the training strategy are made as follows. The ratings of all rated items are set to $1$, representing the clicking behavior. Given an item sequence, our method learns to predict the clicking probability of the last item and ranks the candidate items based on the predicted probabilities \cite{gu2021enhancing}. Accordingly, the cross-entropy is used as the loss function to optimize the model.

\noindent\textbf{Time Complexity Analysis.} The time cost of GNN-DSR lies in the interaction and relational graph aggregation. Given $n$ users, $m$ items, suppose each user directly connects with at most $n_v$ items and $n_s$ social neighbors, and each item directly connects with at most $m_u$ users and $m_c$ correlated items. For interaction aggregation, the major time cost lies in the LSTM and the attention mechanism, as shown in Eq. (\ref{eq:h_iS}) and Eq. (\ref{eq:h_iL}). Since in practice, MLP layers are very small, the time cost for both LSTM and attention are $O(nn_vd^2+mm_ud^2)$. The interaction aggregation is about $O(2nn_vd^2+2mm_ud^2)$. Then, the relational graph aggregation, as shown in Eq. (\ref{eq:h_iN}) and Eq. (\ref{eq:h_jN}), cost $O(nn_sd^2+mm_cd^2)$. The overall time complexity is about $O(n(2n_v+n_s)d^2+m(2m_u+m_c)d^2)$. As $n_v,n_s,m_u,m_c \ll \text{min}\{n,m\}$, the overall time cost is linear to the number of users and items. Therefore, the total time complexity is acceptable in practice.

\section{Experiments}
% In this section, we first introduce the experimental settings, then show the results and make detailed analyses of our method.

\subsection{Experimental Settings}

\subsubsection{Datasets.}

The following three public real-world datasets are used to experimentally evaluate the proposed approach.
\begin{itemize}
  \item Ciao: an online consumer shopping site that records users' ratings of items with timestamps. Items are rated from 1 to 5. Users can also add others to their friend lists and build social connections. 
  % \footnote{\url{https://www.ciao.co.uk}} The Ciao dataset, therefore, contains both user ratings of items and social information. 
  \item Epinions: a famous consumer review site where users can rate items and add social friends to their trust lists. 
  % Epinions\footnote{\url{http://www.epinions.com}} dataset also contains two kinds of information: user-item interaction, where items are rated from 1 to 5 with associated timestamps, as well as the directed trust relationships among users. 
  The Ciao and Epinions datasets \footnote{Dataset available from \url{http://www.cse.msu.edu/~tangjili/trust.html}} have been widely used as benchmark datasets for social recommendations. 
  % \item Douban\footnote{http://www.douban.com} A popular website where users can review items they consume, including but not limited to movies, music, and books. W e used the data in the movie community, obtaining every movie they reviewed along with associated timestamps and their social networks. 
  \item Delicious\footnote{Dataset available from \url{https://grouplens.org/datasets/hetrec-2011/}}: a social bookmarking site where users can store, manage and share web bookmarks. 
  % Users can assign bookmarks to a variety of semantic tags. 
  The Delicious dataset contains bookmarked tags and relationships among users.
\end{itemize}

\begin{table}[t]
  \caption{Statistics of datasets.}
  \centering
  \begin{tabular}{*{4}{c}}
      \hline
      Statistic & Ciao & Epinions & Delicious  \\
      \hline
      \# Users & 2,379 & 22,167  & 1,313   \\
      \# Items & 16,862 & 296,278  & 5,793   \\
      \# Events & 35,990  & 920,073  & 266,190     \\
      \# Social links & 57,544  & 355,813  & 15,328 \\

      \hline
  \end{tabular}
  \label{table:data}
\end{table}

Some statistics for all datasets are provided in Table \ref{table:data}. To evaluate the quality of our method more completely, experiments are conducted both in terms of rating prediction on Ciao and Epinions and item ranking on Delicious. For these three datasets, users who are not socially connected or do not interact, and items that are not interacted with are removed. Modeling these users and items requires a cold-start setting, which is beyond the scope of this work. The cold-start problem will be left to our future work. 

\subsubsection{Baselines.}
To evaluate the performance of our method, three classes of recommenders are used for comparison:
(A) social recommenders, which take into account the social influences; (B) session-based recommenders, which model user sequential interests in sessions; (C) GNN-based recommendation methods, which utilize GNN to capture complex interactions among users and items.

\begin{itemize}
  \item SoReg \cite{ma2011recommender} (A): employs the social network information in a traditional matrix factorization (MF) framework.
  \item SocialMF \cite{jamali2010matrix} (A): incorporates the trust information and its propagation into an MF model. 
  \item DeepSoR \cite{fan2018deep} (A): uses a deep neural network to extract complex and intrinsic non-linear features of each user from social relations for rating prediction.

  \item NARM \cite{li2017neural} (B): applies an attention mechanism to capture the user’s sequential behavior and the user’s main purpose from the current session.
  \item SSRM \cite{guo2019streaming} (B): proposes an MF-based attention model to better understand the uncertainty of user behaviors for the session-based recommendation.
  \item STAMP \cite{liu2018stamp} (B): applies the attention mechanism to better capture users’ short-term interests for the anonymous recommendation. 
  
  \item GraphRec \cite{fan2019graph} (AC): employs a GNN-based framework to model graph data in social recommendations coherently, which can learn user and item representations as well as the heterogeneous strengths of social relations.
  \item DGRec \cite{song2019session} (BC): takes into account both the users’ session-based interests and dynamic social influences. Specifically, this method models context-dependent social influence with a graph-attention neural network. 
  \item GraphRec+ \cite{fan2020graph} (AC): An extension of GraphRec, with additional information of the item-item graph, can learn better user and item representations in social recommendations.
\end{itemize}

For rating prediction, we add rating embeddings to the session-based methods, i.e. methods from class (B) and DGRec, without modifying their model structure and parameter settings.

\subsubsection{Parameter Settings}
% {\color{green}W e implement our method based on pytorch\footnote{https://pytorch.org/}, a well-known python library for deep neural network learning.} 
For each dataset, 80\% is used to be the training set to learn parameters, 10\% as the validation set to tune hyper-parameters, and 10\% as the testing set for the final performance comparison. The dimensions of users and items embedding are set to 128 and the batch size is 256. The lengths of user sequence and item sequence are both truncated to 30. $k=100$ is chosen for building the item correlative graph. The sample neighbors sizes of the social graph and correlative graph are both set to 30. 
% {\color{green}All parameters of the latent factor and weight matrix were initialized by the PyTorch default function that sample from the Gaussian distribution $N(0, 0.01)$, and all biases was set to zeros.} 
The RMSprop \cite{tieleman2012lecture} optimizer is used to train the models with a 0.001 learning rate. The number of hidden layers of the LSTMs is verified at around 4. Dropouts with rates of 0.5 for rating prediction and 0.4 for item ranking are used to avoid overfitting. 
% {\color{green}The parameters of the baseline methods were initialized to those in the corresponding papers.} 

\subsubsection{Evaluation Metrics.}
% To evaluate the recommendation quality, w e conduct experiments both in terms of rating prediction and item ranking. 
For rating prediction, the Mean Absolute Error (MAE) and the Root Mean Square Error (RMSE) \cite{wang2018exploring} are used to evaluate prediction accuracy. Lower values of MAE and RMSE indicate better prediction accuracy. 
% Note that in practice, small improvements in RMSE or MAE terms can have a significant impact on the quality of the top-few recommendations \cite{koren2008factorization}.
For item ranking, Mean Reciprocal Rank at $K$ (MRR@$K$) and Normalized Discounted
Cumulative Gain at $K$ (NDCG@$K$) \cite{chen2021efficient,song2019session} are adopted to evaluate the performance, where $K$ is included $\{10,20\}$. Higher values of MMR@$K$ and NDCG@$K$ indicate better performance of recommender.

\subsection{Quantitative Results}

\begin{table}[t]
  \centering
  \caption{Rating prediction performance of different methods. $\downarrow$ means lower is better. The $p$-value is used to test for statistical significance. * indicates statistically significant improvements $(p<0.01)$ over the best baseline.}
  \begin{tabular}{c|cc|cc}
    \toprule
    \multirow{2}[4]{*}{Models} & \multicolumn{2}{c|}{Ciao} & \multicolumn{2}{c}{Epinions} \\
\cmidrule{2-5}          & RMSE $\downarrow$  & MAE $\downarrow$   & RMSE $\downarrow$  & MAE $\downarrow$ \\
    \midrule
    SoReg  & 1.0848 & 0.8611 & 1.1703 & 0.9119 \\
    SocialMF  & 1.0501 & 0.8270 & 1.1328 & 0.8837 \\
    DeepSoR & 1.0316 & 0.7739 & 1.0972 & 0.8383 \\
    \midrule
    NARM  & 1.0540 & 0.8349 & 1.1050 & 0.8648 \\
    STAMP & 1.0827 & 0.9558 & 1.0829 & 0.8820 \\
    SSRM  & 1.0745 & 0.9211 & 1.0665 & 0.8800 \\
    \midrule
    DGRec & 0.9943 & 0.8029 & 1.0684 & 0.8511 \\
    GraphRec  & 0.9894    & 0.7486    & 1.0673 & 0.8123 \\
    GraphRec+  & 0.9750    & 0.7431     & 1.0627 & 0.8113 \\
    \midrule
    \textbf{GNN-DSR} & \textbf{0.9444*} & \textbf{0.6978*} & \textbf{1.0579*} & \textbf{0.8016*} \\
    \midrule
    $p$-value & $2.07\times10^{-4}$ & $4.05\times10^{-4}$ & $2.89\times10^{-3}$ & $1.44\times10^{-3}$  \\
    \bottomrule
    \end{tabular}%
  \label{tab:rating}%
\end{table}%

\begin{table}[t]
  \centering
  \caption{Item ranking performance of different methods. $\uparrow$ means higher is better. The $p$-value is used to test for statistical significance. * indicates statistically significant improvements $(p<0.01)$ over the best baseline.}
  \begin{tabular}{c|cc|cc}
    \toprule
    \multirow{3}[6]{*}{Models} & \multicolumn{4}{c}{Delicious} \\
\cmidrule{2-5}          & \multicolumn{2}{c|}{MRR@$K$ $\uparrow$} & \multicolumn{2}{c}{NDCG@$K$ $\uparrow$} \\
\cmidrule{2-5}          & $K=10$  & $K=20$  & $K=10$  & $K=20$ \\
    \midrule
    GraphRec  & 0.1437 & 0.1527 & 0.1915 & 0.2243 \\
    GraphRec+  & 0.1568 & 0.1659 & 0.2058 & 0.2392 \\
    \midrule
    NARM  & 0.2011 & 0.2074 & 0.2439 & 0.2680 \\
    STAMP & 0.1993 & 0.2053 & 0.2406 & 0.2626 \\
    SSRM  & 0.1988 & 0.2054 & 0.2392 & 0.2640 \\
    DGRec & 0.2016 & 0.2080 & 0.2669 & 0.2944 \\
    \midrule
    \textbf{GNN-DSR} & \textbf{0.2155*} & \textbf{0.2254*} & \textbf{0.2805*} & \textbf{0.3164*} \\
    \midrule
    $p$-value & $2.56\times10^{-3}$ & $6.54\times10^{-4}$ & $7.64\times10^{-4}$ & $3.57\times10^{-4}$  \\
    \bottomrule
    \end{tabular}%
  \label{tab:topk}%
\end{table}%

The performance of all methods for rating prediction is shown in Table \ref{tab:rating}. 
We have the following observations:
(1) DeepSoR, SoReg and SocialMF all make use of rating and social information. Since DeepSoR is based on the neural network architecture, it performs better than these methods.
(2) Three session-based approaches give slightly better results overall than the social recommenders on Epinions, but do not perform as well as social methods on Ciao. Since the Epinions dataset contains more user-item interactions than Ciao, the user's interests tend to be variable, which are exactly the dynamic interests that session-based methods are good at capturing.
% However, session-based methods only make use of sequential information, whereas the user's interests in Ciao tend to be static, and social methods perform better. 
(3) All three GNN-based recommenders perform better than the previous ones. 
% GraphRec and GraphRec+ perform slightly better than DGRec due to their better integration of rating information, which is an advantage for rating prediction. 
GraphRec+ performs the best for rating prediction among all baselines, as it not only exploits the social influence from users but also considers the correlations between items.
(4) Our proposed method GNN-DSR outperforms all baselines. Compared to the GNN-based methods, our method provides advanced components for integrating user-item temporal interactions and social/correlative graph information. 
% GNN-DSR also takes into account temporal information on both users and items effectively. 
% Details of the $p$-values are shown in \emph{Appendix A}.

The performance for item ranking is shown in Table \ref{tab:topk}.
Since none of the social recommenders in the baselines take into account the user's temporal information, their results of item ranking are all poor. 
% GraphRec+, the best social recommendation baseline in rating prediction, performs much worse than the session-based methods in the item ranking experiment. 
It reflects the importance of temporal information in recommender systems. 
Meanwhile, DGRec performs the best in the baseline models, which demonstrates that dynamically incorporating social information helps to improve the performance of the recommender system. Our method performs the best not only because it captures both dynamic and static information, but also it models the effect of both social and correlative influences.

\subsection{Ablation Study}

To explore the effectiveness of different components in GNN-DSR, an ablation study is conducted on rating prediction by introducing the following variants of our method.
(1) \textbf{GNN-DSR w/o LSTM} only models static representations of users and items.
(2) \textbf{GNN-DSR w/o ATT} only model dynamic representations.
(3) \textbf{GNN-DSR w/o SN} only captures correlative influence from items.
(4) \textbf{GNN-DSR w/o CN} only considers social influence from users.
(5) \textbf{GNN-DSR w/o SC} only uses the information of individual users and items,  neglecting the effect of relational influence.
The results are shown in Table \ref{tab:ablation}.

\begin{table}[t]
  \centering
  \caption{Ablation Study. $\downarrow$ means lower is better.}
  \begin{tabular}{c|cc|cc}
    \toprule
    \multirow{2}[4]{*}{Models} & \multicolumn{2}{c|}{Ciao} & \multicolumn{2}{c}{Epinions} \\
\cmidrule{2-5}          & RMSE $\downarrow$  & MAE $\downarrow$   & RMSE $\downarrow$  & MAE $\downarrow$ \\
    \midrule
    GNN-DSR w/o LSTM & 0.9694 & 0.7456 & 1.0804 & 0.8224 \\
     GNN-DSR w/o ATT & 1.0242 & 0.8350 & 1.0756 & 0.8116 \\
    \midrule
    GNN-DSR w/o SN & 0.9568 & 0.7111 & 1.0621 & 0.8087 \\
    GNN-DSR w/o CN & 0.9496 & 0.7083 & 1.0601 & 0.8069 \\
    GNN-DSR w/o SC & 0.9682 & 0.7303 & 1.0628 & 0.8130 \\
    \midrule
    \textbf{GNN-DSR} & \textbf{0.9444} & \textbf{0.6978} & \textbf{1.0579} & \textbf{0.8016} \\
    \bottomrule
    \end{tabular}%
  \label{tab:ablation}%
\end{table}%

The results show that GNN-DSR w/o ATT gives better prediction than GNN-DSR w/o LSTM on Epinions.
% the short-term representations on Epinions have better predictive capability than the long-term representations.
% , i.e. GNN-DSR w/o ATT gives better prediction than GNN-DSR w/o LSTM. 
It is reasonable since Epinions records a large number of interactions. These interactions may change frequently, so making more use of dynamic data should allow for more effective predictions.
Interestingly, different results are observed on Ciao. 
% Using long-term representations produces more accurate predictions than using short-term ones. 
As Ciao is a small dataset with less variation in interactions, users may tend to have static interactions, as do items. Besides, none of them perform better than the full GNN-DSR.
Thus, capturing dynamic and static information plays different roles in our recommender system, and sensibly combining them will improve the performance.

In both datasets, GNN-DSR w/o CN consistently outperforms GNN-DSR w/o SN, suggesting that the social influence from users has a greater impact on recommendation quality than the correlative influence from items. 
GNN-DSR w/o SC is not as good as GNN-DSR w/o CN and w/o SN, suggesting that combining the influence of relational networks provides a better performance boost to the recommendation system. 
The performance of full GNN-DSR indicates that it is crucial to model the user's current interests and her dynamic social influences, and similarly for items domain.
% GNN-DSR w/o CN, GNN-DSR w/o SN, and GNN-DSR w/o SC all show a significant decrease. Therefore, to achieve good recommendation performance in online communities, it is crucial to model the user's current interests and her dynamic social influences, and similarly for items domain.
% A case study about GNN-DSR and GNN-DSR w/o SC is shown in \emph{Appendix B}.

\begin{figure}[b]
  \centering
  \includegraphics[width=\linewidth]{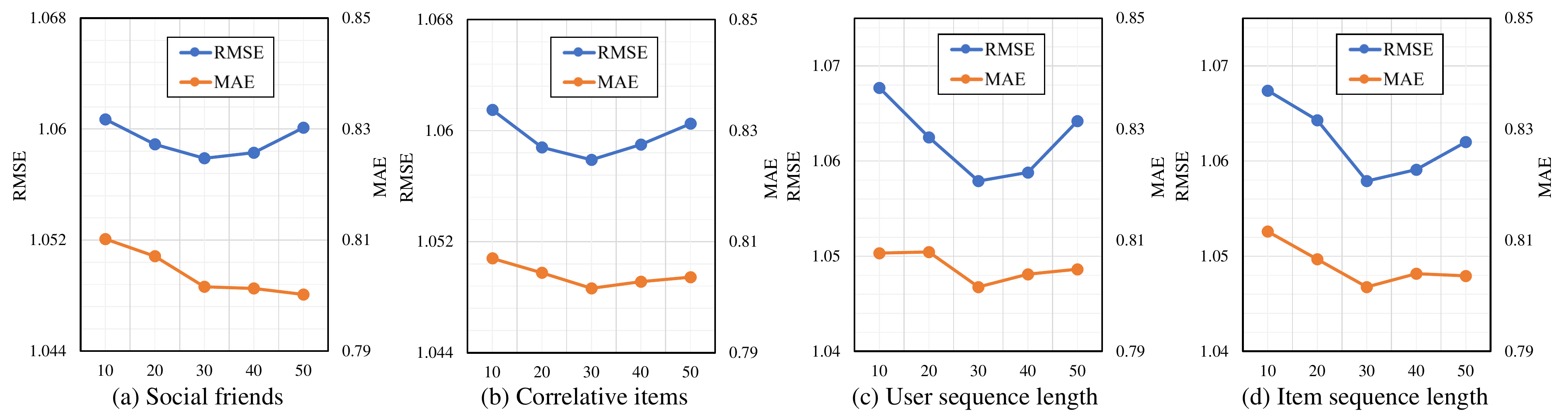}
  \caption{RMSE/MAE of GNN-DSR on Epinions w.r.t different number of friends/items and user/item sequence length.}
  \label{fig:len}
\end{figure}

\subsection{Analysis of Parameters}

% \subsubsection{Key Parameters in User and Item Modeling.}

We investigate the parameter sensitivity of GNN-DSR on some key hyper-parameters, including (a) the number of friends in the social graph, (b) the number of correlative items in the item correlative graph, (c) the user sequence length and (d) the item sequence length. The results are shown in Figure \ref{fig:len} and give us the following findings. 

For (a) and (b), as the number of neighbors increases, both MAE and RMSE decrease at first, indicating that the introduction of relational information is helpful to the proposed method. They tend to rise slowly as the number further increases, except that the MAE of social friends continues to decrease slowly. Maybe a large neighborhood introduces too much noise into the relational graph and RMSE is sensitive to outliers. MAE, on the other hand, may be less affected by these noises so it shows better results with more social influence in social friends. Yet we can still infer from the results of RMSE that the overall performance of the model decreases when too many neighbors are introduced.
%model decreases when too many social friends or correlative items are introduced.
For (c) and (d), as the sequence length grows, there is a significant improvement in performance. The performance decreases when the sequence length is too large. Long sequences mean that the model is biased towards capturing static information, while short sequences are opposite. These experimental results suggest that a reasonable combination of dynamic and static information can help improve model performance.

% \begin{figure}[t]
%   \centering
%   \includegraphics[width=0.6\linewidth]{figure/parameters.pdf}
%   \caption{RMSE/MAE of GNN-DSR on Epinions w.r.t different hyperparameters.}
%   \label{fig:LstmDropout}
% \end{figure}

% \subsubsection{Other Parameters}
% We also investigate the variation in performance of our model under other hyper-parameters, including (a) the number of LSTM layers, and (b) the embedding dimension. The results are shown in Figure \ref{fig:LstmDropout} and give us the following inspirations: 
% (a) The LSTM plays an important role, but if it has too few or too many layers, the model will not perform well. Four layers in our model are performed best. 
% % {\color{green}
% % (b) If the dropout rate is too large, too much randomization will hinder training, while if it is too small, there is no effect of preventing overfitting.} 
% (b) If the embedding size is too small, the model lacks expressiveness. If it is too large, the representations become very sparse, leading to performance degradation. An appropriate size can help to balance the model's expressiveness.
% % {\color{green}
% % (d) As the training sample size increases, the model performance improves overall but the computational cost increases as well. Therefore, appropriate size can help to balance accuracy and complexity.}

\subsection{Case Study}

\begin{figure}[t]
  \centering
  \includegraphics[width=0.88\linewidth]{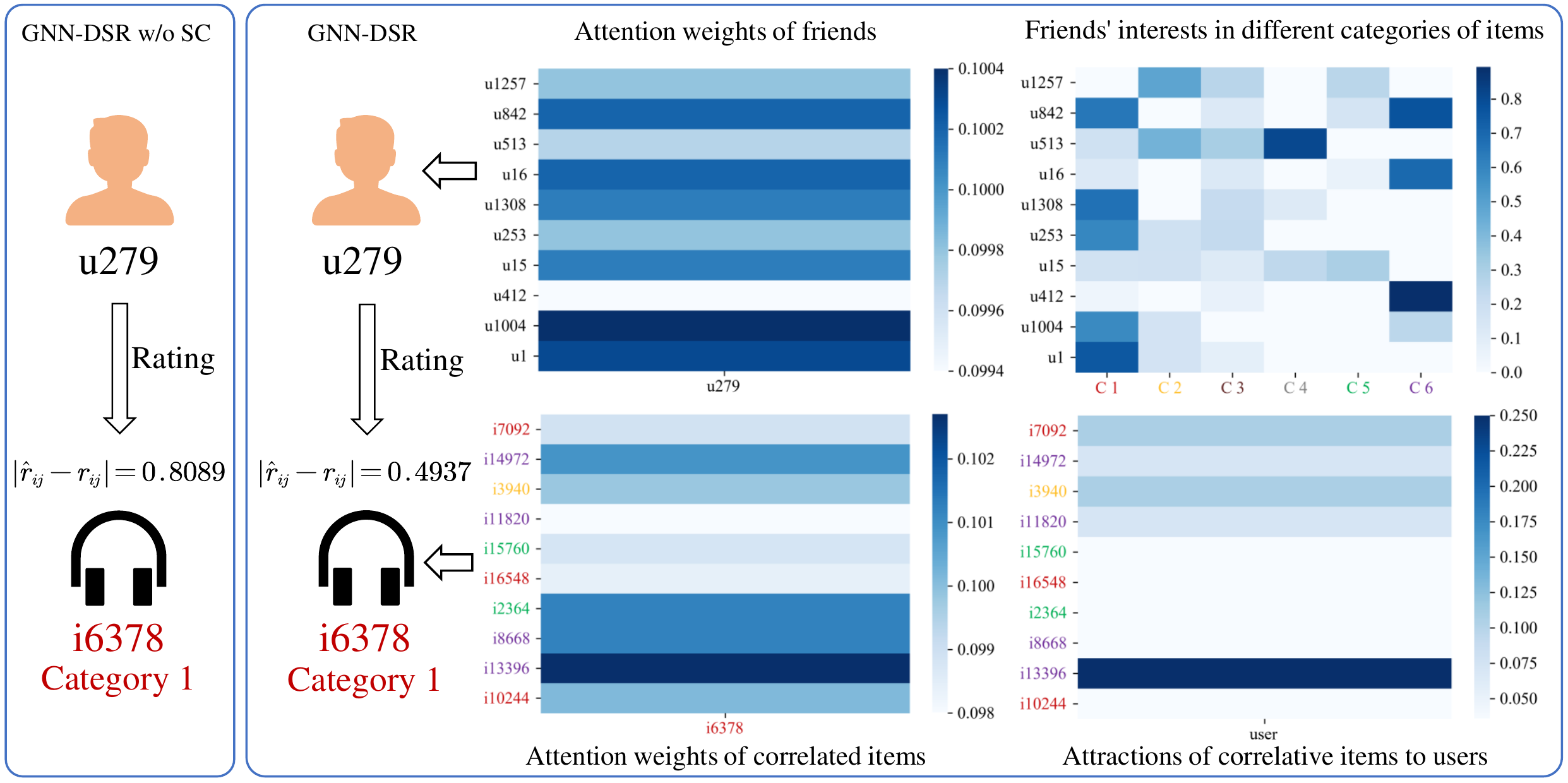}
  \caption{Visualization of the social influence of users and the correlative influence of items in GNN-DSR.}
  \label{fig:Case}
\end{figure}

A major contribution of GNN-DSR is the relational graph aggregation which takes both the social influence of users and the correlative influence of items in learning effective representations for the social recommendation. To visualize the effect of the graph aggregation, a randomly selected user-item pair u279-i6378 is used for conducting a case study on the Ciao dataset, and the results are shown in Figure \ref{fig:Case}. GNN-DSR and GNN-DSR w/o SC are for comparison in this case study.
The number of neighbors for both users and items is kept to 10 for a better presentation. 
There are a total of 6 categories of items on Ciao, distinguished by different colors in Figure \ref{fig:Case}, where C 1 means the items from category 1 and so on. Specifically, i6378 is from category 1. 

On the user side, u279's social friends show different interests when they are confronted with different categories of items. Our method accurately captures the interests of friends and assigns weights to the influence of social friends on the target user through an attention network. In this case, friends who are more interested in the item from categories 1 and 6 seem to have more influence on the target user u279.
On the item side, the items correlated with i6378 mainly come from four categories, where categories 1 and 6 constitute the majority. In this case, item i13396 is the most popular of the correlative items, so it is assigned the largest attention weight, i.e. the target item i6378 is most likely to be influenced by i13396. This result demonstrates the effectiveness of our model in capturing how popular an item is with users and the correlative influence of the related items on the target item.

According to the analysis above, our method can effectively capture social and correlative influences on user and item and efficiently combine them for the recommendation. The result of absolute error, i.e. $|\hat{r}_{ij}-r_{ij}|$ in Figure \ref{fig:Case}, shows that GNN-DSR performs better than GNN-DSR w/o SC, which also verifies the effectiveness of the graph aggregation in our method.

\section{Conclusion}
This paper proposes a GNN-based social recommendation called GNN-DSR. In particular, we consider both dynamic and static representations of users and items effectively. GNN-DSR models the short-term dynamic representations and the long-term static representations in interaction aggregation via RNNs and attention mechanism, respectively. The relational influences from the user social graph or item correlative graph are aggregated via the graph attention mechanism over users' or items' representations in the relational graph aggregation. 
% user modeling and item modeling. For user modeling, GNN-DSR models user interest representations through her short-term dynamic and long-term static interests. Then the user latent factor is modeled by combining interest representations and social influence. For item modeling, our method first constructs an item correlative graph from the raw data for subsequent modeling. GNN-DSR then models dynamic and static attraction representations of the items and combines it with relational influence to model item latent factor. 
% Finally, the user and item latent factors are combined to make a prediction. 
The experimental results of three real-world datasets demonstrate the effectiveness of our method, verifying that items have dynamic and static attraction, and that the correlations among items benefit recommendation.
% We will further incorporate other information such as knowledge graph for better performance.
As to future work, we will further incorporate other side information for solving the cold start problem.
% As to future work, we will focus on how to incorporate other side information such as knowledge graph for making better performance. 
% Since our method runs on existing user-item interactions, it is not friendly for new users or items that are added without much data. Therefore, it would be meaningful to investigate how to incorporate other side information for solving the cold start problem. 

\subsubsection{Acknowledgements.} This work is supported by the National Key R\&D Program of China (2018AAA0101203), and the National Natural Science Foundation of China (62072483).


\bibliographystyle{splncs04}
\bibliography{sample-base}




% LNCS Homepage, \url{http://www.springer.com/lncs}. Last accessed 4
% \end{thebibliography}
\end{document}